\documentclass[twocolumn,superscriptaddress,aps,longbibliography,showpacs,preprintnumbers,amsmath,amssymb,floatfix]{revtex4-1}
\usepackage{titlesec}
\usepackage{amsmath}
\usepackage{bm}
\usepackage{graphicx}
\usepackage[ansinew]{inputenc}
\usepackage{array}
\usepackage{color}
\usepackage{amsxtra}
\usepackage{amstext}
\usepackage{amssymb}
\usepackage{latexsym}
\usepackage{gensymb}
\usepackage{dsfont}
\usepackage{braket}
\usepackage[caption=false]{subfig}
\usepackage[makeroom]{cancel}

\usepackage{balance}

\usepackage{dcolumn}
\usepackage{bm}
\usepackage{bbm}
\usepackage{braket}
\usepackage{mathrsfs}
\usepackage{amstext}
\usepackage{euscript}
\usepackage{txfonts}

\usepackage[dvipsnames]{xcolor}

\usepackage[unicode=true,
 bookmarks=false,
 breaklinks=false,pdfborder={0 0 1},backref=false,colorlinks=true,citecolor=blue]
 {hyperref}

\makeatletter
\@ifundefined{textcolor}{}
{%
 \definecolor{BLACK}{gray}{0}
 \definecolor{WHITE}{gray}{1}
 \definecolor{RED}{rgb}{1,0,0}
 \definecolor{GREEN}{rgb}{0,1,0}
 \definecolor{BLUE}{rgb}{0,0,1}
 \definecolor{CYAN}{cmyk}{1,0,0,0}
 \definecolor{MAGENTA}{cmyk}{0,1,0,0}
 \definecolor{YELLOW}{cmyk}{0,0,1,0}
}

\makeatletter
\renewcommand*\env@matrix[1][*\c@MaxMatrixCols c]{%
  \hskip -\arraycolsep
  \let\@ifnextchar\new@ifnextchar
  \array{#1}}
\makeatother

\newcommand{\cref}[1]{Ref.\,\cite{#1}}

\makeatother

\begin{document}

\title{Quantum eraser from duality--entanglement perspective}

\author{Yusef Maleki}
\email{maleki@physics.tamu.edu}
\affiliation{Department of Physics and Astronomy, Texas A\&M University, 
	College Station, Texas 77843-4242}

\author{Jiru Liu}
\email{ljr1996@physics.tamu.edu}
\affiliation{Department of Physics and Astronomy, Texas A\&M University, 
	College Station, Texas 77843-4242}

\author{M. Suhail Zubairy}
\email{zubairy@physics.tamu.edu}
\affiliation{Department of Physics and Astronomy, Texas A\&M University, 
	College Station, Texas 77843-4242}

\date{\today}

\begin{abstract}
Wave-particle duality is a bizarre feature at the heart of quantum mechanics which refers to the mutually exclusive dual attributes of quantum objects as the wave and the particle. Quantum eraser presents a counterintuitive aspect of the wave-particle duality.
 In this work, we show that quantum eraser can be quantitatively understood in terms of the recently developed duality--entanglement relation. In other words, we show that wave-particle-entanglement triality captures all the physics of the quantum erasure. We find that a controllable partial erasure of the which-path information is attainable, enabling the partial recovery of the quantum interference and extending the scope of the conventional quantum eraser protocols. 
\end{abstract}

\maketitle

\section*{\centering\uppercase\expandafter{\romannumeral1}. Introduction}
Introducing the notion of wave-particle duality, Louis de Broglie put forwards one of the most perplexing concepts of quantum physics in 1923 \cite{deBroglie1923}. Later, this counterintuitive feature was generalazed as the complementarity principle by Niels Bohr \cite{bohr1928}. To be more specific, according to the complementarity principle a quantum object
 has physical properties which are equally real but mutually
exclusive \cite{bohr1928}. To illustrate, considering an interferometry setting, all information contained in a quantum system is captured by both wave and particle nature of the system; however, measuring one of the properties prohibits the other to be observed \cite{bohr1928}. This setting could be understood by examining a single photon that is subjected to an interferometer. In such a discipline, 
the particle nature of the light is captured by our  knowledge  about
the photon path \cite{scully1994,Durr}. In contrast, the wave nature of the light is determined by the visibility of the interference pattern on the screen \cite{scully1994,Durr}. 
\par
The notion of the complementarity principle has been a subject of heated debates since it was first introduced \cite{scully1994,EinsteinLetteSchr1928}; nevertheless, it was not mathematically quantified until 1979, when Wootters and Zurek quantitatively formulated the wave and the particle characteristics of quantum systems \cite{Wootters1979}. This quantification, was later expressed as an explicit inequality $ \mathcal{P}^2+ \mathcal{V}^2\leq 1$ \cite{greenberger1988simultaneous}, where $ \mathcal{P}$ stands for the path information (prior path predictability) of a quantum particle, and $ \mathcal{V}$ stands for the interference pattern, visibility,  addressing the  wavelike behavior of the light  \cite{Glauber1986,Mandel1991,Jaeger1993,Englert1996,rab2017entanglement}. Since then, there has been a great deal of interest in various aspects of the quantum duality \cite{bera2015duality,qureshi2017wave,liu2012relation,huang2013higher,basso2020complete,englert2000quantitative}.
\par
Considering the wave-particle duality in Young double-slit experiment, Scully and Dr{\"u}hl realized a profoundly novel feature that enables recovery of the interference pattern via erasing  the which-path information \cite{scully1982quantum}; a phenomenon that they referred to as "quantum eraser".  This peculiar effect was  later experimented  by Kim, et. al. \cite{kim2000delayed}. As a counterintuitive  aspect of quantum eraser, one may choose to erase
the which-path information even after the photon is registered on the screen, yet restore the interference pattern \cite{aharonov2005time,kim2000delayed}, shedding new lights on the dramatic departure of quantum world from the classical realm \cite{aharonov2005time,kim2000delayed, zubairy2020quantum}.
\\
More recently, it has been discovered that entanglement plays a significant role in the notion of the wave-particle duality.  In fact, it was observed that entanglement controls the duality nature of the quantum systems \cite{jakob2007complementarity,deMelo}. More specifically, taking quantum 
entanglement into the consideration,  the  duality inequality for the double-slit experiment, $\mathcal{P}^2+ \mathcal{V}^2\leq 1$, could be reformulated as the triality relation  $ \mathcal{P}^2+ \mathcal{V}^2+ \mathcal{C}^2= 1$ \cite{jakob2007complementarity,deMelo,Qian2018}. In this equality,  $\mathcal{C}$ is the concurrence, as a measure of two-qubit entanglement \cite{Wootters1998}. This finding exactly quantifies the relation between the complementarity notion of a quantum system and quantum entanglement. Interestingly, a tight connection between the stereographic geometry and the complementarity-entanglement triality of a quantum system was recently established \cite{Maleki}, enabling the full geometric proof and the description of the notion of the complementarity-entanglement triality.
\par
In this work, we show that the complementarity-entanglement triality quantified in terms of $\mathcal{P}$, $\mathcal{V}$ and $\mathcal{C}$ captures all the physics of the quantum eraser protocol, in its general setting. In other words, once the concurrence and the quantitative duality relations are taken into account, we can attain the full description of the quantum eraser protocol, providing insights into the fabric of the quantum complementarity, entanglement and quantum eraser, as the basic concepts that bear the key to the fundamental characteristics of the quantum mechanics, and uncovering their relations. As we show below in this work,  a partial erasure of the which-path information, and hence, partial recovery of the visibility can exactly be realized and characterized using the triality relation. As one main aspect of our work, we extend our analyses to a broader context where the interactions of the system with the external degrees of freedom (i.e., losses to  environment) can also be taken into the consideration, hence, providing a general analysis of the setting.

 %%%%%%%%%%%%%%%%%%%%%%%%%%%%%%%%%%%%%%%%%%%%%%%%%%%%%%%%%%%%%%%%%
\section*{\centering\uppercase\expandafter{\romannumeral2}. Quantum eraser}

\begin{figure}
\begin{center}
\includegraphics[width=7.5cm]{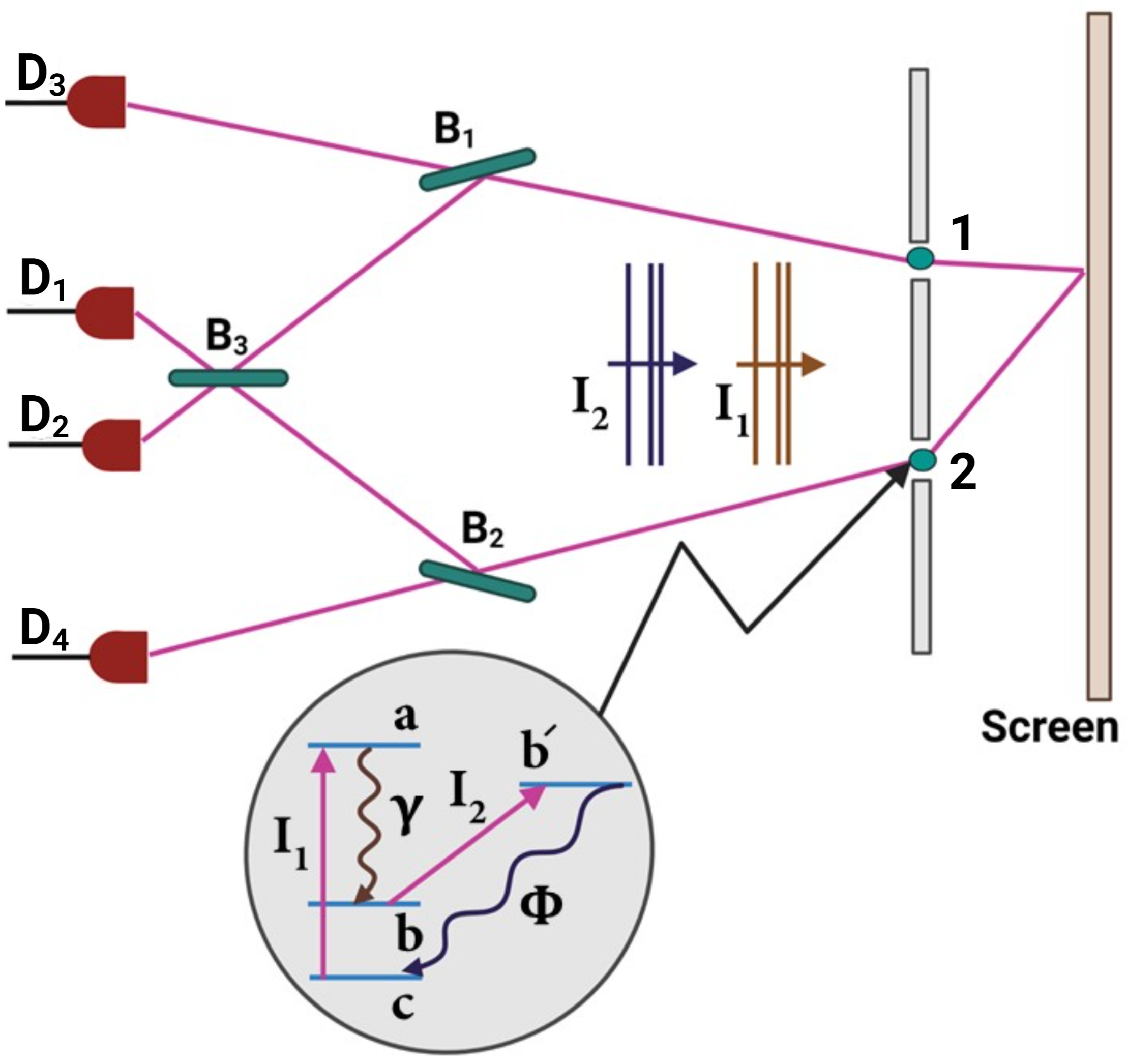}
\caption{Schematics of the quantum eraser experiment, where $D_1-D_4$ are the four single-photon detectors and $B_1-B_3$ are beam-splitters or mirrors. The single-photon pulses $I_1$ and $I_2$, incident on the two atoms, can generate the  $\gamma$ and $\phi$ photon pairs by either atoms at the site 1 or the site 2. The $\phi$ photons proceed to the left while the $\gamma$ photons proceed to the screen on the right.}
\end{center}
\end{figure}

An elegant experimental realization of the quantum eraser strategy was demonstrated in \cite{kim2000delayed}.
The quantum eraser setup that we briefly introduce in this section is investigated by Aharanov and Zubairy in \cite{aharonov2005time} and analyzed in a recent book by Zubairy in
\cite{zubairy2020quantum}, which are akin to the setup in \cite{kim2000delayed}.
The scheme of the setup, for the quantum eraser protocol, is depicted in Fig. 1 \cite{aharonov2005time,zubairy2020quantum}. Accordingly, each pinhole in the double-slit experiment is replaced by a four-level atom. The atoms become excited by the single-photon pulses $I_1$ and $I_2$. The decay of the atoms generates correlated $\gamma$ and $\phi$ photons by either atoms at site 1 or site 2. The energy-level transition diagram and generation of the photon pairs is shown in Fig. 1. In this setting,
each register of $\gamma$ photon on the screen is accompanied by a click on the left-hand-side detectors $D_1-D_4$. Repeating this scattering-detection from the atoms for a large number of times and sorting out the specific registers of the $\gamma$ photon for which a certain detector gets a click, one can recover a destroyed interference pattern by erasing the which-path information. The distribution of the registered $\gamma$ photons on the screen, for the different detection setting of the $\phi$ photons, is depicted in Fig. 2.

To lay out the essence of the quantum eraser in Fig. 1, we start with the state of the photons  emitted by the atoms located at sites 1 and 2 as \cite{aharonov2005time,zubairy2020quantum}
\begin{equation} \label{photon_state0}
    \ket{\Psi}_0=\frac1{\sqrt{2}}\ket{10}_{\gamma}\ket{10}_{\phi}+\frac1{\sqrt{2}}\ket{01}_{\gamma}\ket{01}_{\phi},
\end{equation}
where $\ket{10}_{\gamma}$ stands for the emission of one $\gamma$ photon from site 1 and zero $\gamma$ photon from site 2 (corresponding to the upper path), and similarly $\ket{10}_{\phi}$ stands for the emission of one $\phi$ photon from site 1 and zero $\phi$ photon from site 2 (corresponding to the lower path), etc. 

\begin{figure}
\begin{center} 
\includegraphics[width=7cm,height=5cm]{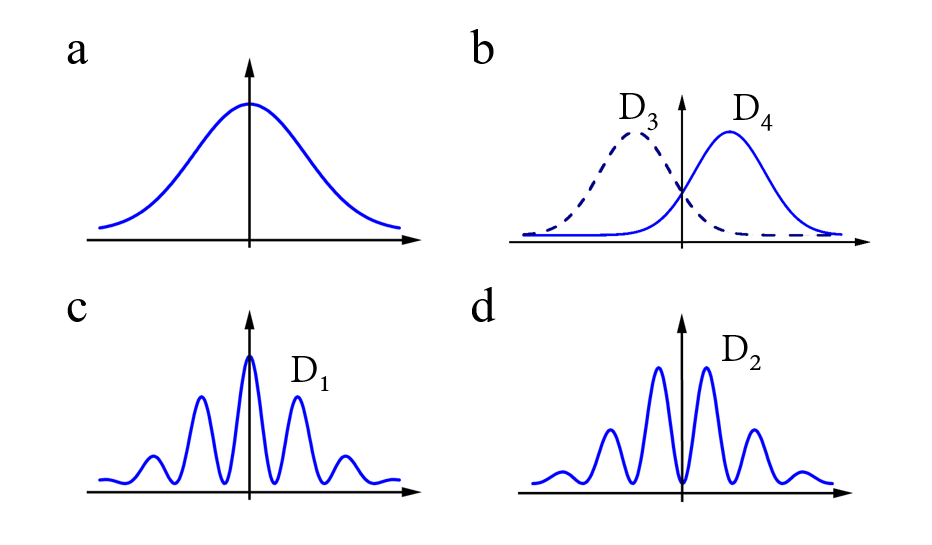}
\caption{Distribution of $\gamma$ photons on the screen. (a) The distribution when no detection are made at $D_1-D_4$.  (b) The distributions for the clicks at $D_3$ and $D_4$, where the full which-path information destroys the interference. The distributions of the $\gamma$ photons for clicks at (c) $D_1$  and (d) $D_2$. In (c) and (d) we do not have the which-path information and interference is recovered.}
\end{center}
\end{figure}
Now, if $B_1$ and $B_2$ are removed (i.e., if they are completely transmitting beam-splitters (BSs)), a $\phi$ photon will be registered   at either $D_3$ or $D_4$. Considering a click at $D_3$, we attain the full which-path information that the photon pair is generated at site 1. Hence, no interference is expected. In this setting, the state $\ket{\Psi}_0$ collapses to $\ket{10}_{\gamma}\ket{10}_{\phi}$. A similar consideration is valid for the register of photon in $D_4$, where the state $\ket{\Psi}_0$ collapses to $\ket{01}_{\gamma}\ket{01}_{\phi}$. In this setting also, as is depicted in  Fig. 2(b), there is no interference. 

In contrast, when we mount  mirrors at $B_1,B_2$ (i.e., if they are perfectly reflecting  BSs), there will be a click at either $D_1$ or $D_2$. The state of the system after passing though the 50:50 beam-splitter $B_3$ becomes 
\begin{equation}
\begin{aligned} \label{photon_state1}
    \ket{\Psi}_1&=\frac12\ket{10}_{\gamma}(\ket{10}_{\phi}+\ket{01}_{\phi})+\frac12\ket{01}_{\gamma}(\ket{01}_{\phi}-\ket{10}_{\phi}) \\
    &=\frac1{2}(\ket{10}_{\gamma}-\ket{01}_{\gamma})\ket{10}_{\phi}+\frac1{2}(\ket{01}_{\gamma}+\ket{10}_{\gamma})\ket{01}_{\phi}.
\end{aligned}
\end{equation}
When $D_1$ clicks the state collapses to $\ket{\gamma_+}$, and alternatively,  when $D_2$ clicks it collapses into $\ket{\gamma_-}$, where
\begin{equation} \label{gamma}
    \ket{\gamma_+}=\frac1{\sqrt{2}}(\ket{10}_{\gamma}+\ket{01}_{\gamma}),\quad \ket{\gamma_-}=\frac1{\sqrt{2}}(\ket{10}_{\gamma}-\ket{01}_{\gamma}).
\end{equation}
We demonstrate in Fig. 2(c,d) the distribution of $\gamma$ photon state $\ket{\gamma_+}$ and $\ket{\gamma_-}$, where the interference pattern from each detection is attainable. While, if no detection is made on the  $\phi$ photons Fig. 2(a) must be attained.

It is worthwhile to mention that an alternative method of attaining which-path information is to setup mirrors at $B_1$ and $B_2$ and remove $B_3$ instead. In this setting, the presence of $B_3$ enables the erasure of the  which-path information and the recovery of the interference patterns.

\section*{\centering\uppercase\expandafter{\romannumeral3}. Duality$-$entanglement relation}
Considering the wave-particle duality scenario, we can encode the existence of a single $\gamma$  photon in path 1 as $\ket{\mathbf{0}}:=\ket{10}$, and the existence of a single $\gamma$ photon  in path 2 as $\ket{\mathbf{1}}:=\ket{01}$. The single photon subjected to the double-slit could be correlated to some external degrees of freedom, akin to the correlation with the $\phi$ photons in the eraser experiment. The entire system of the photon registered on the screen and the correlated system could be expressed as
\begin{equation} \label{whole_system}
\begin{aligned}
     \ket{\Psi}=c_1\ket{\mathbf{0}}\otimes\ket{\phi_1}+c_2\ket{\mathbf{1}}\otimes\ket{\phi_2},
\end{aligned}
\end{equation}

 $c_1$ and $c_2$ being the complex coefficients. The wave-particle duality principle could be formulated as  $\mathcal{P}^2+\mathcal{V}^2\leqslant1$, where $\mathcal{P}$ stands for the path predictability (prior which-path information, that is also referred to as the path distinguishability in some literature \cite{liu2012relation,huang2013higher,Qian2018}) of a quantum object which accounts for the particlelike behavior of the photon, and $\mathcal{V}$ represents the interference pattern visibility,  addressing the  wavelike characteristics of the system  \cite{Jaeger1993,Englert1996}.
\par
The wavelike behavior can be quantified via the fringe visibility on the screen as \cite{Maleki}
\begin{align}
\mathcal{V}=\frac{p_{D}^{max}-p_{D}^{min}}{p_{D}^{max}+p_{D}^{min}}, 
 \end{align}
 where, $p_{D}^{max}$ and $p_{D}^{min}$ are the maximum and minimum probabilities of the photon (photons intensities) registered on the screen, respectively.
 Alternatively, which-path information can be quantified by \cite{Maleki}
\begin{align}
\mathcal{P}=\frac{ | p_{0}-p_{1} |}{ | p_{0}+p_{1} |},
 \end{align}

in which $p_0$ and $p_1$ represent the probabilities of the photon taking the path 1 or path 2, respectively.

The state in Eq. (\ref{whole_system}) recasts a two-qubit state via a proper encoding of the computational basis \cite{Maleki}. With this observation, once the entanglement is taken into the consideration we  attain \cite{jakob2007complementarity,deMelo,Qian2018, Maleki}
\begin{align}
\mathcal{P}^{2}+\mathcal{V}^{2}+\mathcal{C}^{2}=1.
 \label{complementarity2s1}
 \end{align}
 Here,  $\mathcal{C}$ is the concurrence as a measure of two-qubit entanglement \cite{Wootters1998}. The entanglement quantified in the  concurrence framework, is defined through the $R$ matrix as $R=\sqrt{\sqrt{\rho} \bar{\rho}\sqrt{\rho}}$, with $ \bar{\rho}=(\sigma_y \otimes \sigma_y) \rho^{*}  ( \sigma_y \otimes \sigma_y)$. Sorting out the eigenvalues of the  matrix  $R$ in nonincreasing order, one can determine the concurrence as $\mathcal{C}=\text{max}\{0,\lambda_0-\lambda_1-\lambda_2-\lambda_3\}$ \cite{Wootters1998},
where $\lambda_i$s are  the eigenvalues of the matrix $R$   in the deceasing order.

The concurrence of a pure two-qubit state could be determined in terms of the purity of one of the subsystems as $\mathcal{C}^{2}=2[1-\text{Tr}(\rho^2)]$  \cite{Maleki,Wootters2001}, from which we find 
 \begin{align}
\mathcal{P}^{2}+\mathcal{V}^{2}=2\ \text{Tr}(\rho^2)-1.
 \label{complementarity2s2}
 \end{align}

%%%%%%%%%%%%%%%%%%%%%%%%%%%%%%%%%
%%%%%%%%%%%%%%%%%%%%%%%%%%%%%%%%%

\section*{\centering\uppercase\expandafter{\romannumeral4}. Quantum eraser and duality-entanglement relation}

Now, we consider a rather general setting for the quantum eraser protocol and lay out its connection to duality--entanglement relation $\mathcal{P}^{2}+\mathcal{V}^{2}+\mathcal{C}^{2}=1$. As we demonstrate below in this work, the physical picture of the quantum eraser experiment could be fully analyzed via duality--entanglement relation. To this aim, we consider the quantum eraser setup that was discussed earlier. However, to lay out a more general setting, we assume that the probability of the generating photon pairs in site 1 and site 2 are not necessarily symmetric, such that  $|c_1|^2$ and $|c_2|^2$ represent the probability of the pair generation in site 1 and site 2, respectively.
 The state of the entire system could then be written as 
\begin{equation} \label{photon_state2}
    \ket{\Psi}=c_1\ket{10}_{\gamma}\ket{10}_{\phi}+c_2\ket{01}_{\gamma}\ket{01}_{\phi}.
\end{equation}
This state simply reduces to Eq. \eqref{photon_state0} if $c_1=c_2$.
Using this state, we immediately find  the reduced density matrix of the $\gamma$ photon by tracing over the $\phi$ photon, 
from which, we find $ \mathcal{P}=\left||c_1|^2-|c_2|^2 \right|$, $\mathcal{V}=0$, and $\mathcal{C}=2|c_1c_2|$.  Therefore, generation of $\phi$ photon is enough for the vanishing of the fringe visibility. In this case we have $\mathcal{P}^2+ \mathcal{C}^2=1$, as expected. However, detecting $\phi$ photon via $D_{1(2)}$, reduces the $\gamma$ photon state to $\ket{\Psi} = c_1\ket{10}_{\gamma}\pm c_2\ket{01}_{\gamma}$. In this setting, the entanglement  vanishes ($\mathcal{C}=0$), and we  find $ \mathcal{P}=\left||c_1|^2-|c_2|^2 \right|$, $\mathcal{V}=2|c_1c_2|$, leading to $\mathcal{P}^2+ \mathcal{V}^2=1$.

The state in Eq. \eqref{photon_state0}, as well as Eq. \eqref{photon_state2} are pure states, where no interaction with any other degrees of freedom or losses to environment is taken into  consideration. Considering such interactions, the entire state of $\gamma $ and $\phi$ photons will be a mixed state. In a realistic setting the states do not remain pure as the losses into  environment are usually unavoidable \cite{kim2000delayed}. To address this kind of more realistic settings, we need to take  further degrees of freedom into the account. As an interesting  route to incorporate such external degrees of freedom, and more specifically to simulate the effects of decoherence, polarizers could be employed, similar to the approaches taken in \cite{kim2012protecting,kim2009reversing} for simulation of decoherence \cite{escher2011general}. With this strategy in mind, we mount the polarizers $S_1$ and $S_2$ for the photons $\phi_1$ and $\phi_2$, respectively [see Fig. 3]. Considering $\ket{S_1}$ and $\ket{S_2}$ as the polarization states of the site 1 and site 2, the overlap of the polarizations of the $\phi$ photons can be given by $q=\bra{S_1}{S_2}\rangle$, with $0\leqslant |q| \leqslant 1$. Note that when $\ket{S_1}$ and $\ket{S_2}$ are the same, the polarization degree of freedom has no effect on the quantum eraser setup.  The state of the entire system after the $\phi$ photon passes through the polarizer can be written as 
\begin{equation}
    \ket{\Psi}_0=c_1\ket{10}_{\gamma}\ket{10}_{\phi}\ket{S_1}+c_2\ket{01}_{\gamma}\ket{01}_{\phi}\ket{S_2}.
\end{equation}
The density operator in the $\gamma$ and the $\phi$ photons subspace can be attained via $\rho=\text{Tr}_{S}(\ket{\Psi}_0\bra{\Psi}_0)$, such that
\begin{align}
\nonumber
    \rho&=|c_1|^2 \ket{10}_{\gamma} \bra{10}_{\gamma} \otimes \ket{10}_{\phi} \bra{10}_{\phi}
    \\
    \nonumber
    &+|c_2|^2 \ket{01}_{\gamma} \bra{01}_{\gamma} \otimes\ket{01}_{\phi} \bra{01}_{\phi}
    \\
    &+c_1 c_2^* q^* \ket{10}_{\gamma} \bra{01}_{\gamma} \otimes \ket{10}_{\phi} \bra{01}_{\phi} + h.c.
\end{align}
Note that this density matrix recovers the pure state Eq. \eqref{photon_state2} for $q=1$, while it represents a mixed state for $q\neq 1$, and the entire coherence term vanishes when $q=0$.

\begin{figure}
\begin{center}
\includegraphics[width=\columnwidth]{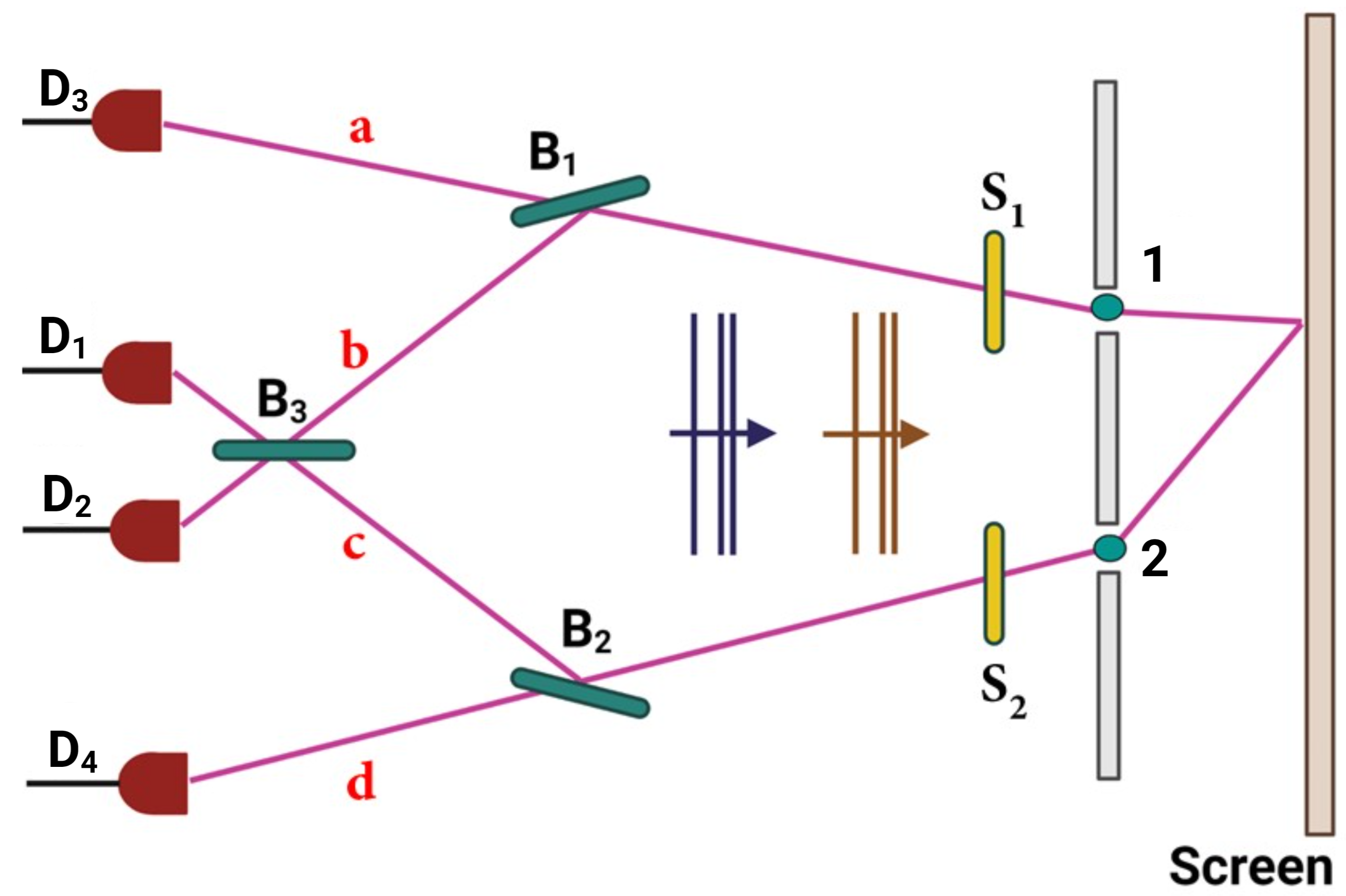}
\caption{Schematics of the generalized quantum eraser experiment, where $D_1-D_4$ are the four single-photon detectors and $B_1-B_3$ are BSs. $S_1$ and $S_2$ are polarizers. The single-photon pulses $I_1$ and $I_2$, incident on the two atoms, assist in generation of the  $\gamma$ and $\phi$ photon pairs by either atoms at site 1 or site 2, as in Fig. 1.}
\end{center}
\end{figure}
\par
Again, in order to lay out a more general setting we assign arbitrary reflectivity and transmitivity to each BS such that the each $B_i$ is characterized by the reflection factor $r_i$ and the transmition factor $t_i$, satisfying $|r_i|^2+|t_i|^2=1$. We can recover the original scheme of the quantum eraser, by setting $|t_1|=|t_2|=1$ for the scenario where $D_3$ and $D_4$ click. In the opposite, we can set $|r_1|=|r_2|=1$ for the scenario where $D_1$ and $D_2$ click. 
\par
With this description, after passing through $B_1$ and $B_2$, the state of the system degenerates into
\begin{align}
\nonumber
    \ket{\Psi}_1
    &=c_1\ket{10}_\gamma[t_1\ket{1000}_\phi+r_1\ket{0100}_\phi]\ket{S_1}
    \\
  & + c_2\ket{01}_\gamma[t_2\ket{0001}_\phi+r_2\ket{0010}_\phi]\ket{S_2},
\end{align}
where $\ket{1000}_\phi$ represents the setting that one $\phi$ photon is on path $a$, and no $\phi$ photon exist on paths $b,c$ and $d$, etc.
\par
After passing though $B_3$, the entire state of the system can be written as

\begin{equation} \label{phi_2}
\begin{aligned}
    \ket{\Psi}_2=&
    c_1 t_{1}\left|\psi_{3}\right\rangle \otimes\ket{1000}_\phi
    + c_2 t_{2}\left|\psi_{4}\right\rangle \otimes\ket{0001}_\phi
\\
   & +N_{1}\left|\psi_{1}\right\rangle \otimes\ket{0010}_\phi+N_{2}\left|\psi_{2}\right\rangle \otimes\ket{0100}_\phi,
\end{aligned}
\end{equation}
where 
$|\psi_{1}\rangle={N_{1}^{-1}}[c_{1} r_{1} r_{3}|10\rangle_{\gamma}|S_{1}\rangle+c_{2} r_{2} t_{3}|01\rangle_{\gamma}\left|S_{2}\right\rangle]$, $|\psi_{2}\rangle={N_{2}^{-1}}[c_{1} r_{1} t_{3}|10\rangle_{\gamma}|S_{1}\rangle-c_{2} r_{2} r_{3}|01\rangle_{\gamma}\left|S_{2}\right\rangle]$, $\left|\psi_{3}\right\rangle=|10\rangle_{\gamma}\left|S_{1}\right\rangle$ and $\left|\psi_{4}\right\rangle=|01\rangle_{\gamma}\left|S_{2}\right\rangle$. Here $N_i$ is determined by the normalization of $\ket{\psi_{i}}$, $i=1,2$.

Once this state is given, we can determine the state of the $\gamma$ photon for the different click of the detectors. 
When $D_3$ and $D_4$ click, the $\gamma$ photon state collapse to $\ket{\gamma}_3=\ket{10}_\gamma$ and $\ket{\gamma}_4=\ket{01}_\gamma$, respectively. While a click on $D_1$ reduces the state of the $\gamma$ photon to $\rho_{\gamma}^{(1)}$ and a click on $D_2$ reduces the state of the $\gamma$ photon to $\rho_{\gamma}^{(2)}$, such that
\begin{equation} \label{gammastate4}
\begin{aligned}
    & \rho_{\gamma}^{(1)}=\frac1{N_1^2}\left[
    \begin{matrix}
    |c_1r_1r_3|^2 & c_2^*r_2^*t_3^*c_1r_1r_3q^*\\
    c_1^*r_1^*r_3^*c_2r_2t_3q& |c_2r_2t_3|^2
    \end{matrix}
    \right] \\
    & \rho_{\gamma}^{(2)}=\frac1{N_2^2}\left[
    \begin{matrix}
    |c_1r_1t_3|^2 & -c_2^*r_2^*r_3^*c_1r_1t_3q^*\\
    -c_1^*r_1^*t_3^*c_2r_2r_3q & |c_2r_2r_3|^2
    \end{matrix}
    \right],
\end{aligned}
\end{equation}
 The probabilities of the click for each detector is given by $p_1= N_1^2=|c_1r_1r_3|^2+|c_2r_2t_3|^2$, $p_2= N_2^2=|c_1r_1t_3|^2+|c_2r_2r_3|^2$, $p_3=|c_1t_1|^2$ and $p_4=|c_2t_2|^2$.

Now, having these relations in hand, we can analyze the different scenarios for the generalized quantum eraser protocol. The click of the detector $D_1$ or $D_2$ contributes to the fringe visibility and the erasing of the which-path information. Without loss of generality, we analyze the click in $D_1$, where the state collapses into $\rho_{\gamma}^{(1)}$. In this case, the visibility, the which-path information and the entanglement are determined as 
\begin{equation} \label{p1v1c1}
\begin{aligned}
    \mathcal{V}_1&=\frac2{N_1^2}|c_1c_2r_1r_2r_3t_3|\cdot|q|, \\
    \mathcal{P}_1&=\frac1{N_1^2}\left||c_1r_1r_3|^2-|c_2r_2t_3|^2\right|, \\
    \mathcal{C}_1&=\frac2{N_1^2}|c_1c_2r_1r_2r_3t_3|\sqrt{1-|q|^2},
\end{aligned}
\end{equation}
 leading to $\mathcal{P}^2_1+\mathcal{V}^2_1+\mathcal{C}^2_1=1$. It is worth mentioning that for calculating the entanglement in the state $\ket{\Psi}_1$, we consider the computational basis of the first subsystem as $|0\rangle\equiv|10\rangle_{\gamma}$ and $|1\rangle\equiv|01\rangle_{\gamma}$, and the computational basis of the second subsystem as $|0\rangle\equiv|S_{1}\rangle$ and $|1\rangle\equiv|S_{2}\rangle$.
 We also note that a similar setting could be obtained for the click in $D_2$. Now, to recover the conventional case of the quantum eraser protocol discussed earlier, we can take $|c_1|=|c_2|$, $|r_1|=|r_2|=1$ and $q=1$; and take $B_3$ as a 50:50 BS. In this scenario, $\mathcal{P}_1$ and $\mathcal{C}_1$ shall be zero, and $\mathcal{V}_1=1$,  with the collapsed state $\ket{\gamma_+}$. Alternatively, the detection of the photons in $D_3$ and $D_4$ always provides $\mathcal{P}_{1(2)}=1$, hence, zero visibility is available. However, considering Eqs. (\ref{phi_2})-(\ref{p1v1c1}), we have provided a much more general setting where even partial visibility and partial which-path information could also  be  quantitatively captured in this setting. Each element in Eq. (\ref{p1v1c1}) can vary from zero to one, depending on the specific parameter choices. 
As a specific setting, the condition $\mathcal{P}_1=0$, provides $|c_1r_1r_3|=|c_2r_2t_3|$. where no which-path information exists. Therefore, the full erasure of the path information can be achieved via specific control of the parameters. In this scenario $ \mathcal{V}_1=|q|$ and $\mathcal{C}_1=\sqrt{1-|q|^2}$. Alternatively, taking $|c_1|=|c_2|$ and $|r_1|=|r_2|=1$  yields $\mathcal{P}_{1}=1$ for $t_{3}=1$ and also for $r_{3}=1$. In this setting, if $B_{3}$ is fully reflective, photons reaching $D_{1}$  always come from site 2, and those reaching $D_{2}$ always come from site $1$, making the full path information available.
As is readily seen from Eq. (\ref{p1v1c1}) the visibility is proportional to $q$. Therefore, the vanishing of the coherence term ($q=0$) destroys the visibility, regardless of which detector clicks. In general, the upper bound to the visibility can be given by  $\mathcal{V} \leqslant q$.
\par
In Fig. 4 we provide different quantum eraser scenarios, where  the duality--entanglement relation provides the full description of the settings. Accordingly, Fig. 4(a) provides the effect of the variation of the polarization degree of freedom. When the two polarizations are orthogonal zero fringe visibility is available. This is due to the fact that the entanglement becomes maximum with respect to the polarization degree of freedom. In Fig. 4(b), the scenario is depicted when $|c_1|$ varies from 0 to 1. In this case, the which-path information can become zero by adjusting the other parameters (when the visibility and the entanglement are maximum), while $|c_1|=0,1$ provides the full which-path information as expected. In Fig. 4(c), the reflection factor of the first BS is varied. As is shown, the maximum of the visibility and the entanglement is attained when the reflectivity is 1. This setting agrees with the which-path information erasing, as the complete reflectivity diminishes the which-path information gain via detector $D_3$. In this setting, the photon is sent to $D_1$ and $D_2$, and hence, the which-path information could be partially erased, and partial visibility could be recovered.  Fig. 4(d) presents the duality-entanglement elements in terms of the reflection factor of the third BS. As is shown, we can control all the elements by changing the reflectively of the third BS, where a partial recovery of the interference and entanglement could be manipulated.

\begin{figure}[h]
\begin{center}
\includegraphics[width=9cm,height=7cm]{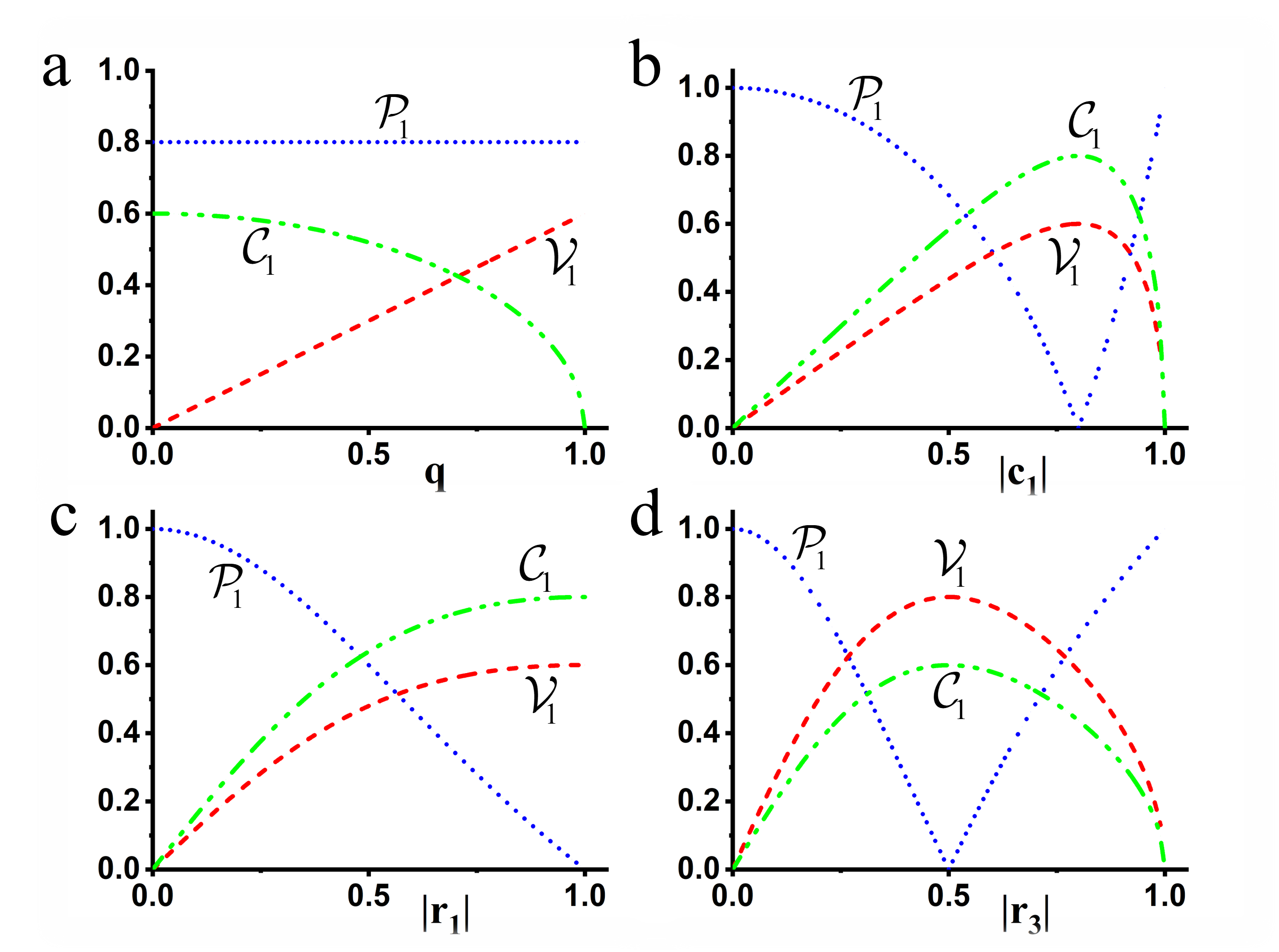}
\caption{ $\mathcal{P}_1$,$\mathcal{V}_1$,$\mathcal{C}_1$ versus (a)  $|q|$, with $|r_1|=|r_2|=1$, $|c_1|^2=0.5$, $|r_3|^2=0.1$. 
(b)  $|c_1|$, with $|r_1|=|r_2|=1$, $|r_3|=0.6$, $|q|=0.6$.
(c)  $|r_1|$, with $|c_1|^2=0.5$, $|r_3|^2=0.5$, $|q|=0.6$. (d)  $|r_3|$, with $|r_1|=|r_2|=1$, $|c_1|^2=0.25$, $|q|=0.6$.}
\end{center}
\end{figure}
%%%%%%%%%%%%%%%%%%%%%%%%%%%%%%%%%%%%%%%%%%%%%%%%%%%%%%%%%%%%%%%%%%%%%%%%%%%%%%%%%%%%%%%%%%%%%%%%%%%%%%%%

In the context of the wave-particle duality, one can also introduce the concept of distinguishability \cite{Englert1996}.
It can be shown that distinguishability is related to predictability and entanglement through $\mathcal{D}_{1}^{2}=\mathcal{P}_{1}^{2}+C_{1}^{2}$ \cite{qureshi2021predictability}.  Considering the state
 \begin{align} \label{d1}
     |\psi_{1}\rangle={N_{1}^{-1}}[c_{1} r_{1} r_{3}|10\rangle_{\gamma}|S_{1}\rangle+c_{2} r_{2} t_{3}|01\rangle_{\gamma}\left|S_{2}\right\rangle],
 \end{align}
and Eq. \eqref{p1v1c1}, we can introduce the distinguishability as
 \begin{align} 
 \mathcal{D}_1=\sqrt{1-\frac4{N_1^4}|c_1c_2r_1r_2r_3t_3|^2|q|^2}.
 \end{align}
In Eq. \eqref{d1}, we can define the probabilities $p_1=|c_{1} r_{1} r_{3}|^2/N_{1}^2$ and $p_2=|c_{2} r_{2} t_{3}|^2/N_{1}^2$, and rewrite the distinguishability as
 \begin{align} 
 \mathcal{D}_1=\sqrt{1-4 p_{1} p_{2}|\langle S_{1}| S_{2}\rangle|^{2}}.
 \end{align}

%%%%%%%%%%%%%%%%%%%%%%%%%%%%%%%%%%%%%%%%%%%%%%%%%%%%%%%%%%%%%%%%%%%%%%%%%%%%%%%%%%%%%%%%%%%%%%%%%%%%%%%%

\section*{\centering\uppercase\expandafter{\romannumeral6}. Summary}

Wave-particle duality is a perplexing feature at the heart of quantum mechanics which refers to the mutually exclusive dual attributes of quantum objects as the wave and the particle.
Quantum eraser presenting a counterintuitive aspect of the wave-particle duality  enables  recovery  of  the  interference pattern  via  erasing  the  which-path  information.
 In this work, we showed that quantum eraser can be quantitatively analyzed in terms of wave-particle-entanglement triality equality. In other words, we showed that wave-particle-entanglement triality captures the entire physics of the quantum erasure. As a part of our analyses, we found that a controllable partial erasure of the which-path information is attainable, enabling the partial recovery of the quantum interference. 
The interesting physics behind the quantum eraser comes from quantum entanglement, which enables the indirect which-path information gain through the detection of the  entangled photon. On the other hand, the duality-entanglement relation is also deeply connected to the quantum entanglement, demonstrating the controlling role of quantum entanglement on the duality. As a result, quantum entanglement is the key to unify these two settings and lay out a general formalism for quantum complementarity.
Our study, illuminating the deep connection between quantum eraser protocols and  wave-particle-entanglement triality, may shed new light on fundamental aspects of quantum interferometry \cite{zubairy2020quantum}, quantum coherence, and quantum steering \cite{maleki2020maximal}.
\section*{Acknowledgements}
%This research is supported by the project NPRP 13S-0205-200258 of the Qatar National Research Fund (QNRF).

This research is supported by the project NPRP 13S-0205-200258 of the Qatar National Research Fund (QNRF).

\bibliography{ref}
\end{document}